# Investigation of Initiation of Gigantic Jets connecting Thunderclouds to the Ionosphere

Lizhu Tong, Kenichi Nanbu, and Hiroshi Fukunishi [1)]

*Institute of Fluid Science, Tohoku University, Sendai, Japan 980-8577*
[1)] *Department of Geophysics, Tohoku University, Sendai, Japan 980-8578*
*E-mail: tong@ifs.tohoku.ac.jp*

The initiation of giant electrical discharges called as "gigantic jets" connecting thunderclouds to the ionosphere is investigated by numerical simulation method in this paper. Using similarity relations, the triggering conditions of streamer formation in laboratory situations are extended to form a criterion of initiation of gigantic jets. The energy source causing a gigantic jet is considered due to the quasi-electrostatic field generated by thunderclouds. The electron dynamics from ionization threshold to streamer initiation are simulated by the Monte Carlo technique. It is found that gigantic jets are initiated at a height of ~18-24 km. This is in agreement with the observations. The method presented in this paper could be also applied to the analysis of the initiation of other discharges such as blue jets and red sprites.

## 1. Introduction

The lightning from thundercloud toward the ground has been widely studied. It has been found, however, that more elusive forms, dubbed elves, sprites with and without sprite halos, and blue jets, flit above the thunderclouds. Recently a new type of lightning, *i.e.*, gigantic jets, was discovered, which linked the top of thunderclouds to the overlying charged atmosphere, known as the ionosphere. Pasko et al. (2002) reported a video recording of a blue jet propagating upwards from a thundercloud to an altitude of ~70 km. The event went across the normal upper limit for blue jets and lower terminal height of sprites. Su et al. (2003) reported their observations of five gigantic jets that spanned the thundercloud top at 16 km and the ionosphere at an elevation of 90 km within half a second. The upper body of gigantic jets was very similar to sprites, but their lower body resembled blue jets. These events are potentially important factors in the model of the earth's electrical and chemical environment. The investigation of these events will help us to understand their contribution to the global electrical circuit (Pasko, 2003).

Observations of gigantic jets are not enough to understand the factors causing the initiation of these events because no associated could-to-ground (CG) lightning discharges were detected in the underlying thunderstorm (Pasko et al., 2002; Su et al., 2003). The conditions triggering gigantic jets are unknown. Finding such conditions is the aim of the present work. Since the extremely-low- frequency (ELF) radio waves associated with gigantic jets were detected in these events and a negative cloud to ionosphere (NCI) discharge could lead to the generation of these ELF waves (Su el al., 2003), here we consider the initiation criteria for gigantic jets which are NCI discharges. The energy source causing gigantic jets is assumed to be due to the quasi-electrostatic field generated by thunderclouds. We calculate the axisymmetrical quasi-electrostatic fields for different thundercloud charges. The criterion of initiation of gigantic jets is proposed using the extension of the triggering conditions of streamer formation in laboratory situations. The estimated initiation heights are in agreement with the observations. Also, we simulate electron dynamics from ionization threshold to streamer initiation and give electron energy distributions at different initiation heights.

**2. Thundercloud charge producing a gigantic jet**

The charge structure of thunderclouds has been studied extensively for many years. MacGorman and Rust (1998) proposed a working hypothesis of the gross charge distribution of the thundercloud, *i.e.*, a positive dipole (positive above negative charge) or a tripole structure with a small lower positive charge plus an upper negative screening layer. Based on these structures, a positive streamer could be initiated if the quasi-electrostatic field above thundercloud exceeds the ionization threshold of air (Pasko et al., 1996; Pasko and George, 2002). However, the hypothesis is inconsistent with the observations of gigantic jets. Pasko et al. (2002) reported that they found an upward transport of negative charges within a gigantic jet. Based on observational results, Su et al. (2003) indicated that gigantic jets might be negative could-to-ionosphere (NCI) discharges. Therefore, here we consider the quasi-electrostatic field above thundercloud generated by thundercloud charge on the basis of a fast accumulation of a negative charge located at the top of thundercloud. The thundercloud charge dynamics can be mathematically represented by (Pasko et al., 1997)

$$Q(t) = Q_0 \frac{\tanh(t/\tau)}{\tanh(1)}, \qquad 0 < t < \tau \tag{1}$$

where $Q_0$ is the magnitude of thundercloud charge and $\tau$ is the duration for accumulation of thundercloud charge. Pasko and George (2002) pointed out that the charge accumulation timescale can in some cases be very fast (fraction of a second). Here we assume the timescale to be ~0.5-1 s. It is noted that the functional variation tanh(·) in eq.(1) is not critical for the physics of the phenomena modeled (Pasko et al., 1997). The thundercloud charge is $Q(t) = \int_V \rho_-(r,z,t) 2\pi r dr dz$, where $V$ represents the whole computational domain. The charge density $\rho_-(r,z,t)$ is assumed to be a Gaussian spatial distribution given by $\rho_-(r,z,t) = \rho(t) e^{-[(z-z_-)^2/a^2 + r^2/b^2]}$, $z_-$ is the mean height of negative thundercloud charges and $\rho(t)$ is the charge density corresponding to $Q(t)$. In this work we set $z_- = 16$ km, $a = 2, 4$, and 6 km, and $b = 2$ km. The charge $Q(t)$ for these three cases of $a$ are shown in Fig. 1 as cases 1, 2, and 3, where $t_{ign1}$, $t_{ign2}$, and $t_{ign3}$ represent the time of initiation of gigantic jets in accordance with three cases of $a = 2, 4$, and 6 km.

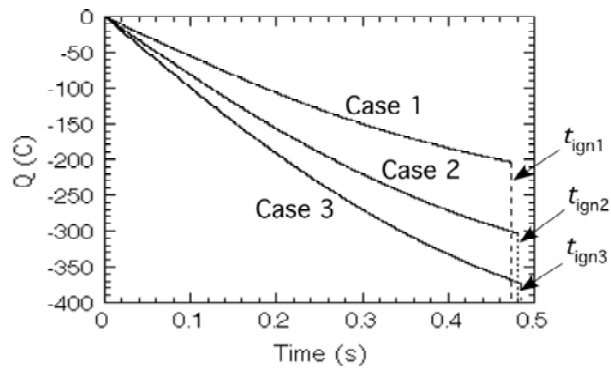

Fig. 1. Temporal evolution of thundercloud charges.

**3. Initiation conditions of gigantic jet**

In order to generate a streamer in air there are three conditions to be fulfilled (Raizer et al., 1998). They are
1) The electric field must exceed the ionization threshold;
2) The initial plasma patch is sufficiently ionized so that the electric field generated by the space charges is comparable with the external field itself;
3) A seed of free electrons is required to start the ionization.

It is known that a minimum electric field strength is required to support stable streamer propagation. Here we pay our attention on negative streamer propagation. The minimum field $-E_c$ required for propagation of negative streamers in air at atmospheric pressure is about $-12.5$ kV/cm (Babaeva and Naidis, 1997), which value has been used to study blue jets (Pasko and George, 2002). The electric field of $E_0 = 3.14 \times 10^6$ V·m$^{-1}$ causes a breakdown of an air gap of 1 cm at atmospheric pressure and produces an effective ionization coefficient $\alpha_{eff} = 1.24 \times 10^3$ m$^{-1}$ (Raizer et al., 1998). The electric field for ionization threshold is estimated by $E_k = E_0 \cdot (N/N_0)$, where $N_0 = 2.688 \times 10^{25}$ m$^{-3}$ and $N$ is the number density of air molecules, taken from US Standard Atmosphere (1976). In this work, we use a simple method to estimate the interaction of charged particles with the external field. The electric field inside a plasma patch generated by electrons and ions is assumed to be equivalent to that caused by the electric dipole such that positive and negative charges locate at the centre of positive ions and electrons, respectively. The electric field in the middle of the electric dipole can be simply given by

$$E_{d/2} = \frac{q}{2\pi\varepsilon_0 (d/2)^2}, \tag{2}$$

where $d$ is the distance of electric dipole, $\varepsilon_0$ is the permittivity of free space, and $q$ is the charge of the electric dipole. We consider that a streamer is initiated if the electric field generated by electrons and ions makes the electric field inside the plasma patch reduce from $E_k$ to $E_c \cdot (N/N_0)$. Therefore, we obtain the minimum number $N_{min}$ of electrons generating a streamer in upper atmosphere by

$$N_{min} = 8.68 \times 10^7 (E_0 - E_c) \cdot (N/N_0) \cdot d^2 \tag{3}$$

The third condition to initiate a streamer can be naturally fulfilled in the domain close to the top of thundercloud if we consider gigantic jets which are negative cloud to ionosphere (NCI) discharges. Our calculation is terminated once the electron number exceeds $N_{min}$. After that, a streamer will be initiated to form a gigantic get.

## 4. Quasi-electrostatic field generated by thundercloud charge

The two-dimensional cylindrical coordinate system $(r,z)$ is used, where the $z$ axis represents the height. The computational domain is $r \leq 60$ km and $0 \leq z \leq 90$ km. We assume that the quasi-electrostatic (QE) field is axisymmetrical. The three boundaries at $z = 0$ km, $z = 90$ km, and $r = 60$ km are assumed to be perfectly conducting. The effect of the artificial boundary at $r = 60$ km on the QE field is considered as small (Pasko et al., 1997).

The continuity equation on the basis of charge conservation law is (Tong et al., 2004)

$$\frac{\partial \rho}{\partial t} + \nabla\sigma \cdot \mathbf{E} + \rho\sigma/\varepsilon_0 = 0, \tag{4}$$

where $\rho$ is the charge density, $\sigma$ is the conductivity, and $t$ is the time. $\mathbf{E}$ is electrostatic field governed by

$$\nabla \cdot \mathbf{E} = (\rho + \rho_s)/\varepsilon_0, \tag{5}$$

where $\rho_s$ is the thundercloud source charge density, *i.e.*, $\rho_s = \rho_-$ in this work. Before the ionization threshold is reached, electron conductivity below 60 km height is low. Therefore the total conductivity σ is dominated by ion conductivity, taken by $\sigma = 5 \times 10^{-14} e^{z/6km}$ S/m (Dejnakarintra and Park, 1974). The ordinary finite difference method (Potter, 1973) is used to solve eq. (4). The Fourier transform method and Thomas algorithm (Hockney and Eastwood, 1988) are used to solve eq.(5).

## 5. Monte Carlo method simulating electron dynamics

The one-dimensional Monte Carlo simulation starts at the time that electric field arrives at ionization threshold. Several hundred, or thousand, electrons are assumed as emitted from the top of thunderclouds. The equation of motion for the electrons is simulated by the modified Verlet method (Ueda, 1990)

$$z(t + \Delta t) = z(t) + v_z(t)\Delta t + F_z(t)(\Delta t)^2 / 2m , \qquad (6)$$

$$v_z(t + \Delta t) = v_z(t) + [F_z(t + \Delta t) + F_z(t)]\Delta t / 2m , \qquad (7)$$

where $z$ and $v_z$ are the axial components of position and velocity of electrons, respectively, $m$ is electron mass, $\Delta t$ is time step, and $F_z$ is the force acting on electrons given by $F_z = q_e E_z$. Here $q_e$ is electronic charge and $E_z$ is the axial component of electric field. For a mixture of $N_2$ and $O_2$, we consider 33 types of electron-molecule collisions. In addition to elastic, rotational, and vibrational excitations, we consider electronic excitations of $N_2(A^3\Sigma_u^+)$, $N_2(B^3\Pi_g)$, $N_2(W^3\Delta_u)$, $N_2(B'^3\Sigma_u^-)$, $N_2(a'^1\Sigma_u^-)$, $N_2(a^1\Pi_g)$, $N_2(w^1\Delta_u)$, $N_2(C^3\Pi_u)$, $N_2(E^3\Sigma_g^+)$, $N_2(a''^1\Sigma_g^+)$, and 13 eV, ionizations of $N_2^+(X^2\Sigma_g^+)$, $N_2^+(A^2\Pi_u)$, and $N_2^+(B^2\Sigma_u^+)$ for $N_2$, and electronic excitations of $O_2(a^1\Delta_g)$, $O_2(b^1\Sigma_g^+)$, $O_2(c^1\Sigma_u^-)$, 6.0 eV, 8.4 eV, and 10 eV, ionizations of $O_2^+(X^2\Pi_g)$, $O_2^+(a^4\Pi_u)$, $O_2^+(A^2\Pi_u)$, $O_2^+(b^4\Sigma_g^-)$, and $O_2^+(^4\Sigma, ^2\Sigma_s)$, dissociative attachment, and dissociative excitation for $O_2$. The corresponding cross sections for these reactions are taken from Rees (1989), Rapp and Briglia (1965), and Phelps (1985). For the estimation of electron-molecule collisions, Nanbu (1994) proposed a simple method to simultaneously determine whether an electron collides and which collisional event occurs in the case of collision.

By the method, we obtain the collision probability of the *i*th electron for the *k*th collisional type in $\triangle t$

$$P_k = n_n \sigma_k(\varepsilon_i) v_i \Delta t , \qquad (k = 1, 2, \dots, K) \qquad (8)$$

where, $\varepsilon_i$ and $v_i$ are the energy and speed of the *i*th electron, $\sigma_k(\varepsilon_i)$ is the cross section of the *i*th electron for the *k*th type collision, $n_n$ is the density of neutral gas, $N_2$ or $O_2$, and $K$ is the total number of collisional types, *i.e.*, $K = 33$ in the present work. The total probability that an electron collides with a molecule in $\triangle t$ is

$$P_T = \sum_{k=1}^{K} P_k \qquad (9)$$

Equation (9) is written again as

$$1 = P_T + (1 - P_T) = \sum_{k=1}^{K} \left[ P_k + \left( \frac{1}{K} - P_k \right) \right] \qquad (10)$$

Based on eq. (10), the method to sample a collisional event is shown in Fig. 2. The unit length is divided into *K* equal intervals and each interval is divided into two. We call a uniform random number *U* (0 < *U* < 1). The integral part of *KU*+1 is the number of the *k*th interval in which *U* lies. The left part of the *k*th interval is regarded as $1/K - P_k$ and its right part is $P_k$. If *U* lies in $P_k$, the *k*th event occurs; otherwise the particle does not collide. Regarding the method for determining the post-collisional velocity of electron, please see Nanbu (2000).

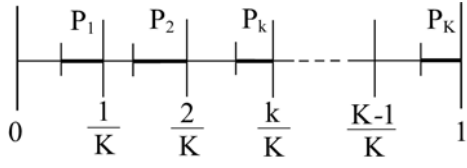

Fig. 2. The method to sample a collisional event.

## 6. Results and discussion

We simulate the evolution of electric field and charge density during the accumulation of thundercloud charge for cases 1-3. Figure 3 shows the distribution of electric fields along $z$-axis for cases 1-3. The electric field generated by thundercloud charges just exceeds the critical field $E_k$ for ionization. The first initiation condition of a gigantic jet is fulfilled. Table I gives the thundercloud charge $Q$, the height $H$ of ionization threshold, and the time $t_{ioz}$ corresponding to Fig. 3. $t_{ioz}$ is the duration from the accumulation of thundercloud charge to such a case that ionization threshold of air is reached, i.e., $E \geq E_k$ at $t = t_{ioz}$.

Table I. $Q$ (C), $H$ (km), $t_{ioz}$ (s) corresponding to ionization threshold

| Cases | $Q$ | $H$ | $t_{ioz}$ |
| --- | --- | --- | --- |
| Case 1 | 203.57 | 18.63 | 0.47 |
| Case 2 | 303.63 | 20.74 | 0.48 |
| Case 3 | 373.27 | 23.55 | 0.48 |

As seen in Table I, the heights of 18.63-23.55 km for ionization threshold are in agreement with the observations of Su et al. (2003). They reported that an apparent emerging point of the gigantic jets started from the heights of 18 km, 22 km, and 24 km. The thundercloud charges of 200-400 C have been used to study sprites and blue jets (Pasko et al, 1996, 1997). The magnitudes of charges shown in Table I could be reached by the accumulation of thundercloud.

In the Monte Carlo simulation we consider that the neutral gases are composed of 80% $N_2$ and 20% $O_2$. We assume that about one thousand electrons are emitted at the heights of ionization threshold. Then the electrons are accelerated and collide with neutral air molecules. We follow electron dynamics until the electron number reaches $N_{min}$ given by eq. (3). The time $t_{str}$ to initiate a streamer from ionization threshold is 1.94, 2.53, and 3.68 μs for cases 1-3, respectively. The initiation time of gigantic jets is estimated as $t_{ign} = t_{ioz} + t_{str}$. Figure 4 gives the distribution of electrons at the time of initiation of gigantic jets. The coordinate $z$ is counted from the ionization height $H$ shown in Table I. For the ionization at a lower height, the motion of electrons is spatially more restricted due to high atmospheric pressure. With the increase of ionization height, electrons start to disperse, such is seen in case 3 in Fig. 4.

In this work we examined the electron energy distribution. The electron energy distribution $E(\varepsilon)$ at the time of initiation of gigantic jets is given in Fig. 5. As seen in Fig. 5, the EEDs for cases 1-3 have a similar distribution, deviating from Maxwellian. The distribution has a high energy tail in which the energies of only a few electrons are over the ionization thresholds of gases, i.e., 15.58 eV for $N_2$ and 12.1 eV for $O_2$ (Tong et al., 2004). The calculation shows that the largest electron energy appears around ~16-20 eV. The average electron energy is ~4.97 eV, which is consistent with the previous research (~5 eV) of sprites (Pasko et al., 1997).

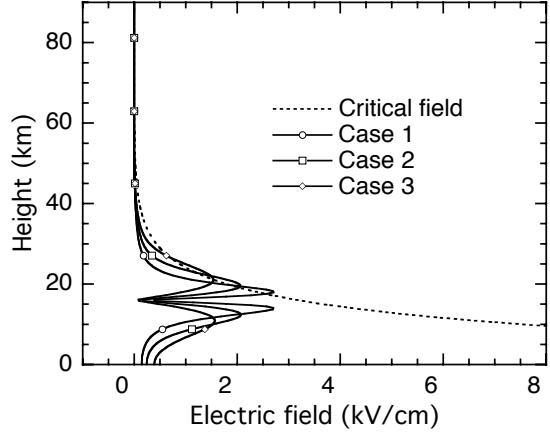

Fig. 3. Electric field on the axis ($r = 0$) compared with critical field $E_k$ for ionization.

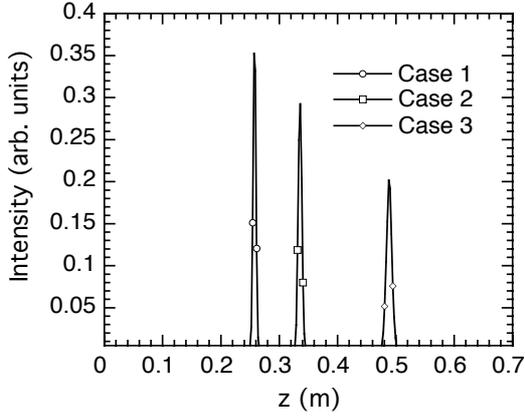

Fig. 4. The distribution of electrons started from the ionization heights at the time of initiation of gigantic jets.

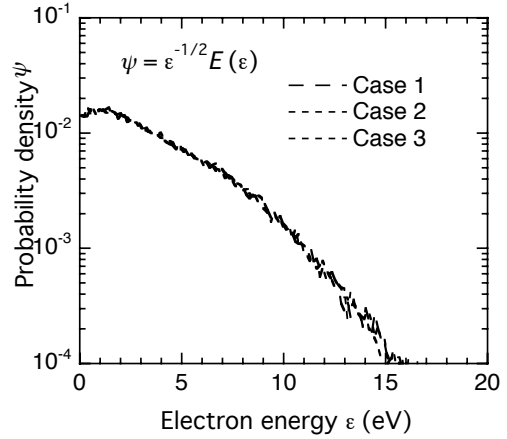

Fig. 5. Electron energy distribution at the time of initiation of gigantic jets.

## 7. Conclusion

We investigate the initiation of gigantic jets connecting thunderclouds to the ionosphere by numerical simulation method. A criterion determining the initiation of gigantic jets is proposed on the basis of the triggering conditions of streamer formation in laboratory situations. Based on the present model, we found that the gigantic jets appear at ~18-24 km height, which is consistent with the observations. The electron energy distribution during the initiation of gigantic jets deviates from Maxwellian distribution. The average electron energy is ~4.97 eV and the largest electron energy reaches ~16-20 eV.

The present model presents the first step in studying gigantic jets. In reality, the gigantic jets possess a number of branches generated by different charged plasma channels. Studies of multiple streamers in the upper atmosphere is a much more complicated problem, which will be our next aim. These studies will make us have a better understanding for gigantic jets, a new giant electrical discharge phenomenon. This is not only interesting for basic science, but also important for solving aeronautics problems in the future.